\documentclass[aps,prb,preprint,floatfix,superscriptaddress,showpacs]{revtex4-1}
\usepackage{graphicx}% Include figure files
\usepackage{dcolumn}% Align table columns on decimal point
\usepackage{multirow}
\usepackage{amsmath,bm}% bold math
\usepackage{xcolor}
\usepackage{pdfcomment}
\usepackage[normalem]{ulem}
\begin{document}

%\listoftodos
%\newpage

\title{A ``non-dynamical'' way of describing room-temperature paramagnetic manganese oxide}
%\title{Electronic structure calculation of paramagnetic Mott insulator MnO\\ based on static density functional theory}

\author{Sangmoon Yoon}
\affiliation{Department of Materials Science and Engineering,
             Seoul National University, Seoul, 08826, Korea}
\affiliation{Department of Physics and
             Research Institute for Basic Sciences,
             Kyung Hee University, Seoul, 02447, Korea}
            
\author{Seoung-Hun Kang}
\altaffiliation{Present address: Korea Institute for Advanced Study
             (KIAS), Seoul, 02455, Korea.}
\affiliation{Department of Physics and
             Research Institute for Basic Sciences,
             Kyung Hee University, Seoul, 02447, Korea}
            
\author{Sangmin Lee}
\affiliation{Department of Materials Science and Engineering,
             Seoul National University, Seoul, 08826, Korea}
\affiliation{Department of Physics and
             Research Institute for Basic Sciences,
             Kyung Hee University, Seoul, 02447, Korea}
            
\author{Kuntae Kim}
\affiliation{Department of Materials Science and Engineering,
             Seoul National University, Seoul, 08826, Korea}
             
\author{Jeong-Pil Song}
\altaffiliation{Present address: Department of Physics, 
             University of Arizona, Tucson, AZ, 85721, USA.}
\affiliation{Department of Physics and
             Research Institute for Basic Sciences,
             Kyung Hee University, Seoul, 02447, Korea}
             
\author{Miyoung Kim}
\email[Corresponding author. E-mail: ]{mkim@snu.ac.kr}
\affiliation{Department of Materials Science and Engineering,
             Seoul National University, Seoul, 08826, Korea}

\author{Young-Kyun Kwon}
\email[Corresponding author. E-mail: ]{ykkwon@khu.ac.kr}
\affiliation{Department of Physics and
             Research Institute for Basic Sciences,
             Kyung Hee University, Seoul, 02447, Korea}
%\affiliation{Korea Institute for Advanced Study (KIAS), Seoul, 02455, Korea}

\date{2019-05-06}

%---------------------------------------------------------------------
\begin{abstract}
We present a new approach based on the static density functional theory (DFT) to describe paramagnetic manganese oxides, representative paramagnetic Mott insulators. We appended the spin noncollinearity and the canonical ensemble to the magnetic sampling method (MSM), which is one of the supercell approaches based on disordered local moment model. The combination of the noncollinear MSM (NCMSM) with DFT$+U$ represents a highly favorable computational method called NCMSM$+U$ to accurately determine the paramagnetic properties of MnO with moderate numerical cost. The effects of electron correlations and spin noncollinearity on the properties of MnO were also investigated. We found that the spin noncollinearity plays an important role in determining the detailed electronic profile and precise energetics of paramagnetic MnO. Our results illustrate that the NCMSM$+U$ approach may be used for insulating materials as an alternative to the \textit{ab initio} framework of dynamic mean field theory based on DFT in the simulation of the room-temperature paramagnetic properties.
\end{abstract}
%---------------------------------------------------------------------

\pacs{
71.27.+a, 
% Strongly correlated electron systems; heavy fermions
71.15.Mb,
% Density functional theory, local density approximation, gradient and other corrections
75.20.-g,
%Diamagnetism, paramagnetism, and superparamagnetism
75.30.Et
%   Exchange and superexchange interactions (see also 71.70.Gm Exchange interactions)
%\towho[inline]{SHK}{Verify the PACS numbers}
%\mycomment[inline]{SHK}{Done}
}

% \keywords{schwarzite, density functional theory, electronic structure}


\maketitle

%---------------------------------------------------------------------

\section{Introduction}

% entire page. There is no need to do explicit centering.
\begin{figure*}[t]
\includegraphics[width=1.0\textwidth]{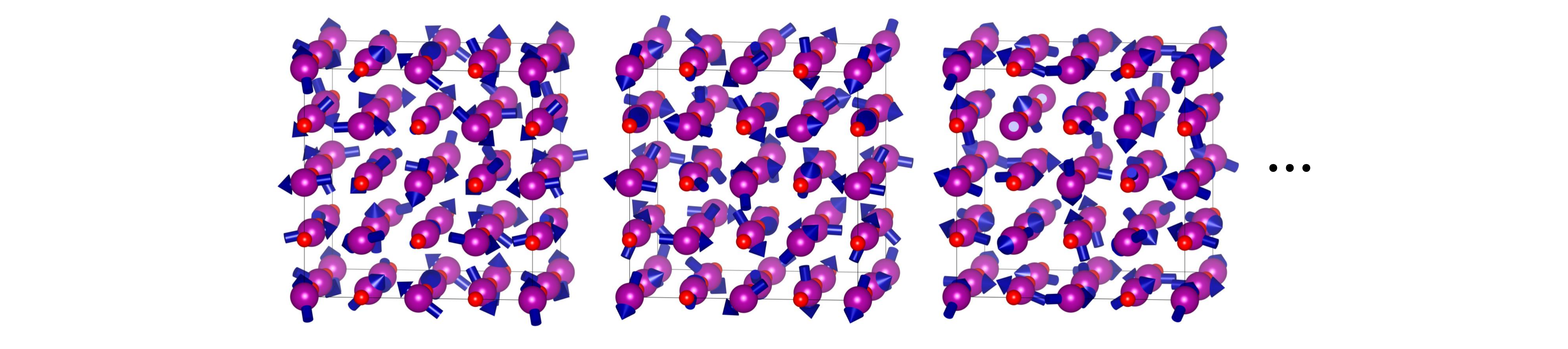}
\caption{(Color online) 
Schematic illustration of the noncollinear magnetic sampling method (NCMSM) for the paramagnetic Mott insulating phase of MnO. The detailed explanations are given in the text.
\label{Fig1}}
\end{figure*}
%---------------------------------------------------------------------

% entire page. There is no need to do explicit centering.
\begin{figure}[t]
\includegraphics[width=1.0\columnwidth]{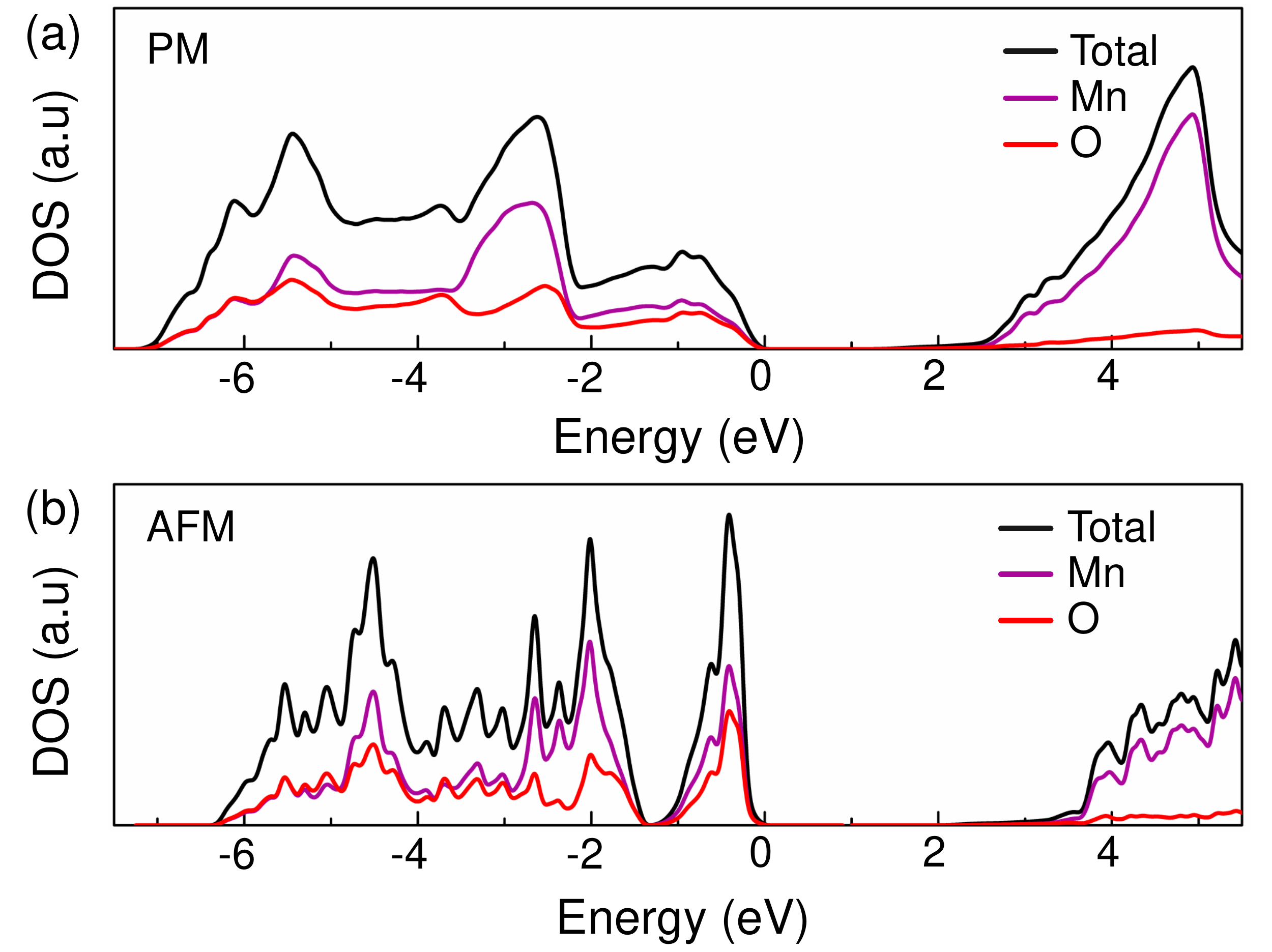}
\caption{(Color online) 
DOSs of (a) paramagnetic (PM) and (b) antiferromagnetic (AFM) MnO phases calculated with $U=4$~eV. The PM DOS was obtained using our NCMSM$+U$ approach. The black solid line represents the total DOS; the purple and red solid lines indicate the partial DOSs of Mn and O, respectively.
\label{Fig2}}
\end{figure}
%---------------------------------------------------------------------

Manganese oxides have received a lot of attention as being among the most promising materials for applications in acquisition and storage of sustainable energy, for example, in catalysts, supercapacitors, and lithium-ion batteries. They have a potential to afford various oxidation or reduction reactions as manganese has a multivalent character, and nanosynthesis and surface treatment techniques are also being rapidly developed to support such capabilities. Furthermore, manganese is an earth abundant and eco-friendly element. Thus, attempts are underway to use nano-sized manganese oxides in energy-related devices~\cite{{JACS1},{EES1},{EES2},{EES3},{JACS2},{AC1},{SR1},{JACS3},{NL1},{AM1}, {AFM1},{JACS4},{JMC2},{SCI1},{JMC1},{SR2},{SR3},{ACSN1},{ACSN2},{NL2},{NL3},{ACSN3}}. For more effective and efficient development of such devices, it is highly required to investigate systematically and theoretically the room-temperature properties of nanostructured manganese oxides.

Manganese oxides are usually paramagnetic Mott insulators under ambient conditions, exhibiting Mott insulating state despite of spin fluctuations because of the strong correlations of $d$ electrons. Thus, for the theoretical study of manganese oxides, it is a primary task to confirm the Mott insulating states. Unfortunately, density functional theory (DFT) calculations  fail consistently to describe such Mott insulating state because of its spurious self interactions. In fact, this failure had been one of the central issues in DFT studies of the past decades. Now, it is partially resolved through the Coulomb-interaction corrections of localized electrons such as DFT$+U$~\cite{{PRB7},{PRB8}}, hybrid functional~\cite{{JCP1},{JCP2}}, GW~\cite{PRL2} or self-interaction-correction~\cite{PRB9}. Manganese oxides have also been intensively studied by adopting these approaches~\cite{{PRB1},{PRB12},{PRB13},{CM1},{CM2},{JPCC1},{CM3},{JPCC2}}. However, all these extended DFT methods require the materials to be in spin- or orbital-ordered state, because the electron correlations are still corrected under the static mean field level.

The consistent description of paramagnetic Mott insulators became possible after the development of DFT plus dynamic mean field theory (DFT+DMFT)~\cite{{PRL1},{PRB6}}, where both strong correlations and spin fluctuations are spontaneously involved through the frequency-dependent self-energy. It was, for example, used to understand the paramagnetic Mott insulator to metal transition of MnO~\cite{NATM1}. Moreover, this method has provided a new direction in the research of many strongly correlated materials, such as heavy fermion systems~\cite{{SCI2},{PRB15}} and high-temperature superconductors~\cite{{PRL3},{PRB14}}. However, DFT+DMFT is not suitable for the material design of paramagnetic Mott insulators, because its application to surfaces, interfaces, defects, or various other configurations with large super cells is nearly unfeasible. Meanwhile, the disordered local moment (DLM) based DFT$+U$ calculations have also been recently discussed~\cite{PRB16}. The DLM-based approach involves the effects of spin fluctuations at the static limit. It has successfully described high-temperature properties of some specific correlated systems, for example, several transition-metal oxides and nitrides~\cite{{PRB16},{CrN0},{CrN3},{CrN4},{NJP1},{PRL8}}. The DLM approach has been implemented in DFT$+U$ in two different manners. One employs the Korringa-Kohn-Rostoker (KKR) Green's function method with coherent potential approximation (CPA), and the other uses supercells that imitate magnetic disorder. The KKR-CPA method can calculate the electronic structure of disordered systems efficiently, but the expansion beyond the bulk calculation is limited because of the spherical approximation for one-electron potential. On the other hand, the supercell approach does not have such restrictions, but it generally requires hundreds of atoms to minimize the spurious interactions from the periodic repetition of magnetic disorder. That is, both approaches have advantages and disadvantages. As the computing power has increased rapidly and the need for material research has grown, the supercell approach has naturally gained more attentions, and the related methodologies have been further developed.~\cite{{PRB16},{CrN0},{CrN3},{CrN4}}.

% entire page. There is no need to do explicit centering.
\begin{figure*}[t]
\includegraphics[width=1\textwidth]{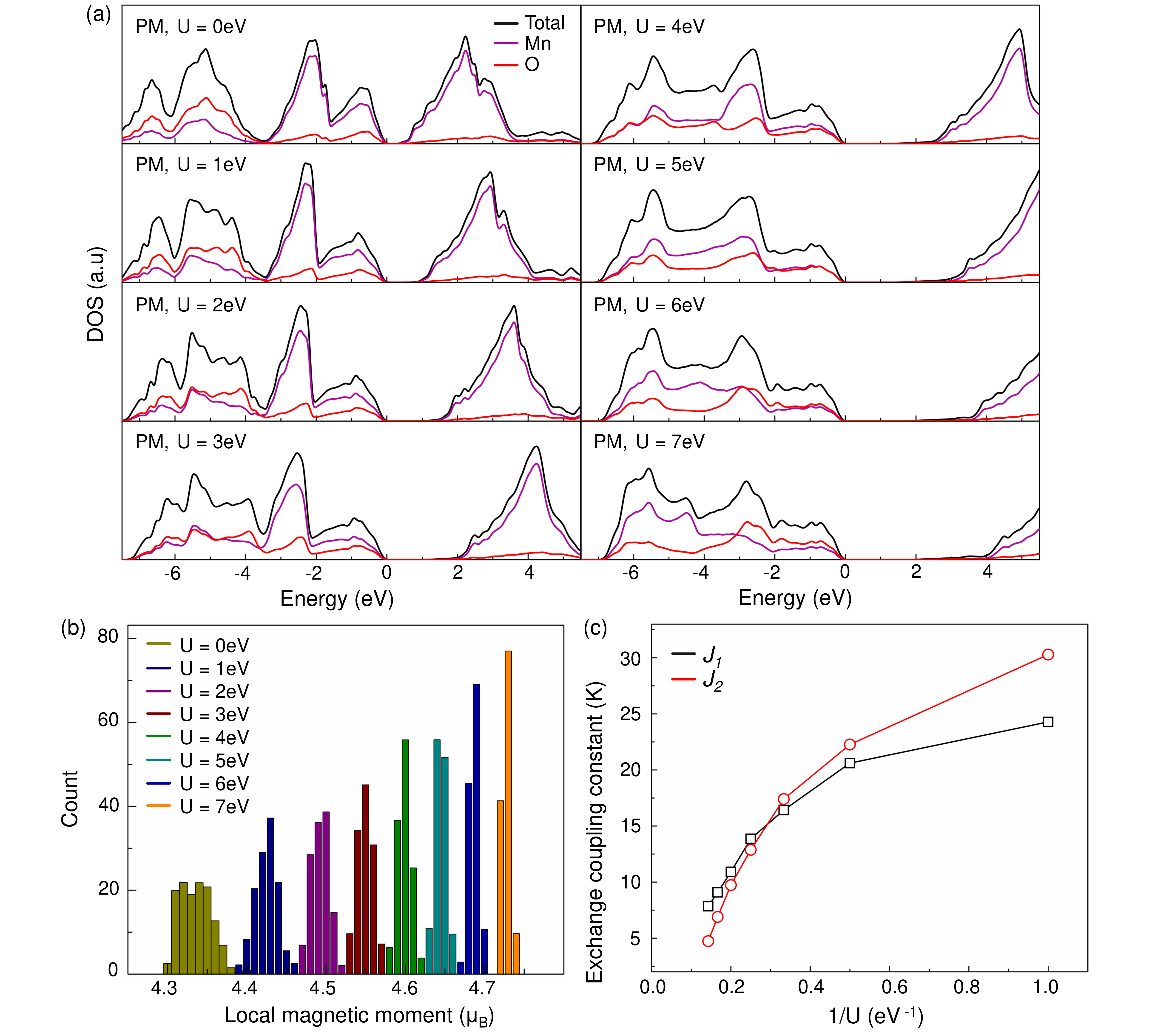}
\caption{(Color online) 
The effects of electron correlations on the electronic and magnetic properties of paramagnetic MnO. (a) DOSs of paramagnetic MnO with different values of on-site Coulomb interaction $U$. (b) The distribution of local magnetic moments of the paramagnetic MnO for different values of $U$, which are represented by different colors. (c) Dependence of Heisenberg exchange couplings $J_1$ (depicted by squares with black line) and $J_2$ (circles with red line) with the nearest and next-nearest neighbors, respectively on the on-site Coulomb interaction $U$.
\label{Fig3}}
\end{figure*}
%---------------------------------------------------------------------

Magnetic disorder in a paramagnetic state can be characterized by the spin-spin correlation function, defined as
\begin{equation}
\label{equation1}
  \langle\Phi_{\alpha}\rangle=\frac{1}{N}\sum_{i,j\in\alpha}
      \mathbf{S}_i\cdot\mathbf{S}_j,
\end{equation}
where $\alpha$ and $N$ refer to an index of the specific coordination shell and the total number of spins, respectively. $\mathbf{S}_i$ is the spin angular momentum of the $i$-th spin. For ideal paramagnetic disordered materials, $\langle\Phi_{\alpha}\rangle$ in Eq.~(\ref{equation1}) is zero regardless of the type of coordination shell. Supercells of the DLM model should, in principle, involve the nature of paramagnetic disorder, but it is practically hard to construct such supercells because of their finite size. For a realistic supercell-based DLM calculation, two approximate approaches were proposed~\cite{PRB16}. The first approach employs a special quasirandom structure, where the spin-spin correlation function for the finite coordination shell is only vanished. The other approach uses randomly disordered magnetic structures and describes the paramagnetic disorder as their average, which is called the magnetic sampling method (MSM). In MSM, it is supposed that the spin-spin correlation function is averaged out as the number of magnetic samples increases. 

Generally, in Mott physics, quasiparticles can emerge with renormalized masses and finite life times, displaying a three-peak spectral structure. However, for insulating phase, spectral weight transfers from the quasiparticle peak to the Hubbard bands, and thus quasiparticles eventually disappear at the Mott transition and dynamical charge correlations become drastically slow. Hence, it is expected that the DLM-based DFT$+U$ calculation could describe the physical properties of paramagnetic Mott insulators if the effects of magnetic disorders and static self-energies are included adequately. We, therefore, further improved the MSM approach to be accurate and computationally scalable for paramagnetic Mott insulators. In specific, we revised the MSM approach to consider the spin noncollinearity and to take canonical ensemble average. We named this approach ``noncollinear MSM based on DFT plus $U$'' (NCMSM$+U$); it is illustrated in Fig.~\ref{Fig1}. The motivation and importance of the inclusion of spin noncollinearity will be discussed in the result section.

To verify the validity and efficiency of our NCMSM$+U$ approach, we selected paramagnetic MnO, which is not only a representative example of paramagnetic manganese oxides where quasiparticles are suppressed, but also an important material for real applications. It has been directly utilized for oxygen evolution reaction catalysts~\cite{{AC1},{SR1},{JACS3}} or for lithium ion batteries~\cite{{JMC1},{SR2},{SR3}}. Thus, the accurate estimation of various room-temperature properties of paramagnetic MnO is an important task for rational device design in real applications. In this paper, we report that the NCMSM$+U$ approach yields various properties of paramagnetic MnO comparable to those calculated from DFT+DMFT. We also investigate the effects of electron correlations and spin noncollinearity on those properties, and discuss the application of our NCMSM$+U$ method to the paramagnetic materials with superexchange interactions. 

%---------------------------------------------------------------------
\begin{table}[t]
%\centering
%\begin{ruledtabural}
\begin{tabular}{c|cccccc} 
  \hline\hline
       & NCMSM$+U$   & Expt. & Expt. & Expt. & Expt. & Expt. \\ 
       & (This work) & (Ref.~56) & (Ref.~57) & (Ref.~58) & (Ref.~59) & (Ref.~49)\\ \hline
  $J_{1}$ & 13.9 & 10 & 8.9 & 7.2 & 8.7 & - \\ 
  $J_{2}$ & 12.8 & 11 & 10.3 & 3.4 & 10.4 & - \\ 
  $J_{2}/J_{1}$ & 0.92 & 1.10 & 1.16 & 0.47 & 1.20 & 1.49 \\
  \hline\hline
\end{tabular}
%\end{ruledtabural}
\caption{Exchange coupling constants $J_{1}$, $J_{2}$ (in K) and their ratio $J_{2}/J_{1}$, and the comparison with other experimental values.
\label{Tab1}}
\end{table}
%---------------------------------------------------------------------

% 2. Computational 
\section{Computational details and methodology}
\label{Computational}

We constructed magnetically noncollinearlly disordered supercells for NCMSM using two random number generators, which independently determine the azimuthal and polar angle of each magnetic moment. We set the initial magnitude of each magnetic moment to an experimental value of 4.58~$\mu_\mathrm{B}$~\cite{PRB25}, where $\mu_\mathrm{B}$ is the Bohr magneton, with a constraint of the total magnetic moment to be zero. (See the Supporting Information.) Although a larger supercell would provide better results, such as the effects of the spin correlations, we used the $2\times2\times2$ supercell with 64 atoms (32 manganese atoms and 32 oxygen atoms) for NCMSM due to limited computational resources. A series of DFT calculations were conducted using the Vienna \textit{ab initio} simulation package (VASP) code~\cite{PRB24}. The Perdew-Burke-Ernzerhof plus Hubbard correction (PBE$+U$) was used for the exchange-correlation functional~\cite{PRL6}, in which the double-counting interactions were corrected in the fully localized limit (FLL)~\cite{{PRB8},{ArXiv1}}. Here, it should be discerned that the parameter required in noncollinear model is the on-site Coulomb interaction $U$ whereas that required in collinear model is the effective on-site Coulomb interaction $U_{\textrm{eff}}$~\cite{ArXiv1}. A plane wave basis set with a cutoff energy of 500~eV was used to expand the electronic wave functions, and the valence electrons were described using the projector-augmented wave potentials. The lattice constants in all the cases were fixed to the experimental value of 4.4315~\AA~\cite{PRB25} to prevent the artificial deformation that would be induced by the magnetic disorder. The $\Gamma$-centered $4\times4\times4$ Monkhorst-Pack $k$-point grid was used for sampling the Brillouin zone. Antiferromagnetic phase was also investigated using the same parameters as used in magnetic disordered phase to clearly see the effects of magnetic disorder, without considering a significant structural distortion the antiferromagnetic MnO~\cite{PRB19} undergoes. 

We not only implemented the spin noncollinearity into the MSM approach, but also introduced the sample averaging scheme in the canonical ensemble, in which a new configuration was contributed to the physical properties depending on the weight of each magnetic sample proportional to the Boltzmann factor. Then, any physical quantity in the paramagnetic phase, $X^\mathrm{PM}$, was evaluated by an ensemble average defined by
\[
X^{\mathrm{PM}}=\frac{\sum_iX_ie^{-E_i/k_\mathrm{B}T}}{\mathcal{Z}},
\]
where $\mathcal{Z}$ is a partition function defined by $\sum_{i}e^{-E_{i}/k_\mathrm{B}T}$. $X_i$ and $E_i$ are the specific physical quantity and the energy of the $i$-th microstate, respectively, and $k_\mathrm{B}$ is the Boltzmann constant. In particular, the total energy, density of states (DOS) and local magnetic moment distributions were all computed by taking ensemble average. We used seven different disordered structures for the ensemble average at $T=300$~K. The probability of finding each microstate was in the range of 12 to 16~\% at room temperature, implying that there was no certain specific microstate dominant in the ensemble of paramagnetic MnO (see Fig. 1S in the Supplementary Information).

% 3. Results
\section{Results and discussion}
\label{Results}
\subsection{Electronic Structures of Paramagnetic MnO}
\label{Results1}

% entire page. There is no need to do explicit centering.
\begin{figure}[t]
\includegraphics[width=0.7\columnwidth]{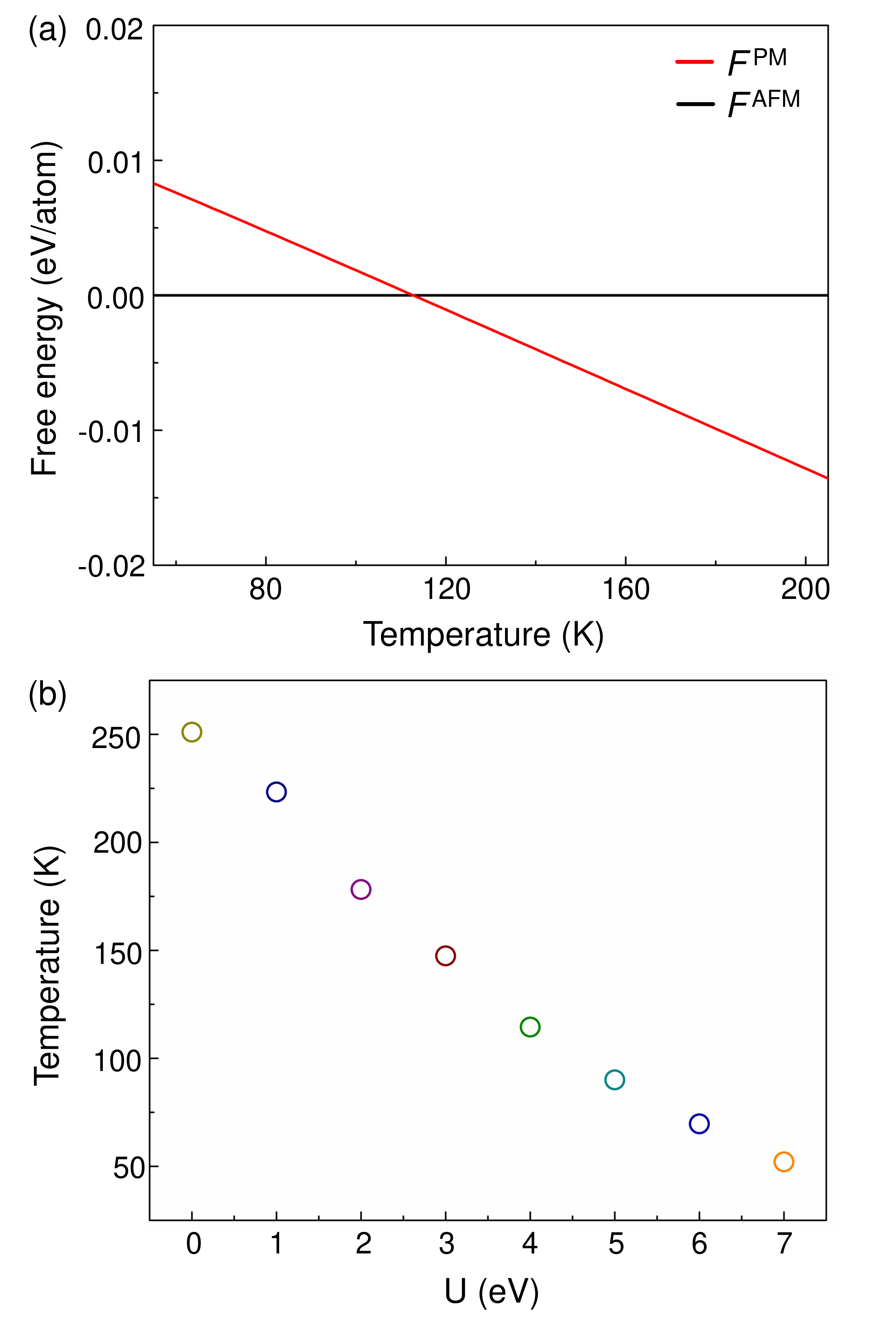}
\caption{(Color online) 
(a) Per-atom Helmholtz free energy of paramagnetic (PM, red solid line) MnO relative to that of antiferromagnetic (AFM, black solid line) MnO, which is set to zero, obtained with $U=4$~eV. Temperature at the crossover point is approximately 114~K, which corresponds to the N\'eel temperature. The N\'eel temperature of MnO is experimentally known to be 118~K. (b) The crossover temperatures corresponding the N\'eel temperature estimated with various values of $U$.
\label{Fig4}}
\end{figure}
%---------------------------------------------------------------------

Figure~\ref{Fig2}(a) shows the DOS of the paramagnetic MnO determined using the NCMSM$+U$ calculation. We chose an on-site Coulomb interaction to be $U=4$~eV. This value enabled us to obtain the results that agree well with the experimental measurements, which will be shown in the following subsections. The paramagnetic MnO DOS calculated by the NCMSM$+U$ approach reveals features distinct from that of its antiferromagnetic counterpart shown in Fig.~\ref{Fig2}(b). It displays a wider band width and a narrower band gap in the paramagnetic state compared to those in the antiferromagnetic state. Moreover, unlike the DOS of the antiferromagnetic phase, the DOS of the paramagnetic phase exhibits delocalized electronic characteristics in the valence band, as seen in Fig.~\ref{Fig2}(a). Such behavior gives rise to crucial consequences for electron correlations in the strongly-correlated MnO. It is noteworthy that, for the paramagnetic MnO, our numerical results are in good agreement with previous estimates from DFT+DMFT~\cite{NATM1} and the experimental XPS spectra~\cite{NATM1}. This result can be considered a step-up improvement in the static DFT$+U$ based calculations for the strongly correlated systems. In usual DFT calculations, nonmagnetic states have been used to mimic paramagnetic states. We checked that the nonmagnetic MnO is metallic even in the presence of the Hubbard corrections (see Fig.~2S in Supplementary Information). This is consistent with the previous knowledge that DFT$+U$ requires spin- or orbital-ordered states. Our NCMSM$+U$ method provides an alternative way of achieving what the DFT+DMFT detects.

\subsection{Effects of Electron Correlations}
\label{Results2}

We investigated the effects of strong electron correlations on the electronic structure of the paramagnetic MnO by adjusting the Hubbard parameter $U$ from 0 to 7~eV. Figure~\ref{Fig3}(a) displays the total and partial DOSs of the paramagnetic MnO for various $U$ values. The majority-spin (minority-spin) bands shift down (up) with $U$. Here, the valence and conduction bands of MnO correspond one-to-one with the majority- and minority-spin bands, since the Mn ions have $d_{5}$ high-spin configuration. In addition, the band shift is accompanied with the band gap increase. The band gap here is of the charge-transfer type. At a relatively high $U$, the system enters a charge transfer state with the minority-band shift from the Mott-Hubbard phase, as seen in Fig.~\ref{Fig3}(a). Thus, our results suggest that the paramagnetic MnO is a mixed type of Mott insulator with the band gap of 2.6 eV.

We also investigated the electron correlation effect on the magnetic properties of paramagnetic MnO. We produced canonical ensemble composed of individual microstates or paramagnetic configurations, which were generated through the self-consistent procedure performed using our NCMSM$+U$ approach. The initial magnetic moments were chosen to be 4.58~$\mu_\mathrm{B}$, corresponding to the experimental value. Figure~\ref{Fig3}(b) shows the distribution of the local magnetic moments calculated with the different values of $U$. It is clearly shown that the choice of $U$ values determines the distribution of the local magnetic moments in paramagnetic states. The mean value of each magnetic moments distribution obtained for given $U$ increases monotonically from 4.33 to 4.73~$\mu_\mathrm{B}$ with $U$. This result is consistent with the previous knowledge that the on-site Coulomb repulsion enhances the spin- and orbital-polarization~\cite{PRB26}. On the other hand, the distribution width decreases with increasing $U$, that is, the local magnetic moments are widely (narrowly) distributed for small (large) values of U. From our NCMSM$+U$ calculation, we determined the on-site Coulomb parameter of $U=4$~eV, because this value yields the mean magnetic moment value of $4.59~\mu_\mathrm{B}$, which is the closest value to the experimental value.

The interatomic superexchange couplings are directly related to the on-site Coulomb interactions. We extract the exchange coupling constants by mapping the energies of disordered states onto the Heisenberg Hamiltonian, obtained in the limit of $U\rightarrow\infty$,
\begin{equation}
\label{equation3}
\mathcal{H}=J_1\sum_{\langle{i,j}\rangle}\mathbf{S}_i\cdot\mathbf{S}_j+J_2\sum_{\langle\langle{i,j}\rangle\rangle}\mathbf{S}_i\cdot\mathbf{S}_j,
\end{equation}
where $\langle{i,j}\rangle$ and $\langle\langle{i,j}\rangle\rangle$ denote the first- and second-nearest neighboring Mn-Mn pairs, and $J_1$ and $J_2$ are their corresponding exchange coupling constants. Figure~\ref{Fig3}(c) shows the exchange coupling constants $J_1$ and $J_2$ in the units (K) of the absolute temperature, as a function of $1/U$, which were fitted over data obtained from seven different disordered structures. Both $J_1$ and $J_2$ decrease monotonically with increasing $U$, and are all positive, meaning that both the first- and second-nearest exchange interactions are antiferromagnetic regardless of the strength of $U$ interactions. Our fitted $J_1$ and $J_2$ values are almost linear in the large $U$ region or for small $1/U$  as shown in Fig.~\ref{Fig3}(c), which is consistent with the well-known linear expression $J=2t^2/U$ for the superexchange coupling in the strong $U$ regime. Our NCMSM$+U$ with $U=4$~eV yielded $J_{1}$=13.9~K and $J_{2}$=12.8~K, resulting in their ratio of $J_2/J_1\approx0.92$, which is in good agreement with experimental results as shown in Table~\ref{Tab1} ~\cite{{PR3},{SSC1},{PRB20},{JPCS1},{JPCS2},{PRL8}}

\subsection{Estimation of N\'eel Temperature}
\label{Results3}

To estimate the N\'eel temperature $T_\mathrm{N}$, which is the transition temperature from the antiferromagnetic ($T<T_\mathrm{N}$) to the paramagnetic states ($T>T_\mathrm{N}$), we evaluated the Helmholtz free energies $F^\mathrm{AFM}$ and $F^\mathrm{PM}$ of antiferromagnetic and paramagnetic states in MnO as a function of temperature $T$. In general, the Helmholtz free energy $F$ of a system, given by $F=E-TS$, can be determined by computing the internal energy $E$, and the entropy $S$ at a given temperature $T$.  The entropy $S(T)$ of a magnetic system under the mean-field approximation can be expressed as
\begin{equation}
\label{equation4}
S(T) = k_\mathrm{B}\ln[M(T)+1],
\end{equation}
where $M(T)$ indicates the mean value of local-magnetic-moment distribution. For the antiferromagnetic case, $M(T)$ should be zero at the mean-field level, and thus its Helmholtz free energy $F^\mathrm{AFM}$ is simply given in terms of the internal energy
\[
F^\mathrm{AFM}(T)=E^\mathrm{AFM},
\]
independent of temperature. On the other hand, $F^\mathrm{PM}$, the free energy of the paramagnetic phase, can be expressed as
\[
F^\mathrm{PM}(T)=E^\mathrm{PM}(T)-k_\mathrm{B}T\ln[M(T)+1],
\]
where $E^\mathrm{PM}(T)$ is the ensemble average of the internal energies calculated for seven different spin configurations. The temperature dependences of both the internal energy and the mean local magnetic moment were obtained through the Boltzmann factors used in the ensemble average. Figure~\ref{Fig4}(a) shows the evaluated Helmholtz free energies of both antiferromagnetic and paramagnetic states of MnO as a function of temperature $T$ for $U=4$~eV. The paramagnetic free energy, which is higher than the antiferromagnetic free energy at low temperatures, crosses $F^\mathrm{AFM}$ at $T\approx114$~K and becomes lower than the antiferromagnetic counterpart above the crossover temperature. The crossover temperature corresponding to the N\'eel temperature was estimated to be $T_N=114$~K, which is surprisingly close to the experimental value of 118~K~\cite{PRL5}.

In addition, we estimated the crossover temperatures for several different values of $U$. The corresponding free energies at each $U$ are presented in Fig.~3S of Supplementary Information. As shown in Fig.~\ref{Fig4}(b), the N\'eel temperature decreases from 250~K to 50~K with increasing $U$. This trend is closely related to the weakening of the exchange interactions. For comparison, we also estimated the N\'eel temperature at PBE level. The result was more than twice higher than that obtained with our NCMSM$+U$ approach, which is consistent with the previous PBE-based result~\cite{PRB16}.

\subsection{Importance of Spin Noncollinearity in Paramagnetic MnO}
\label{Results4}

% entire page. There is no need to do explicit centering.
\begin{figure}[t]
\includegraphics[width=0.7\columnwidth]{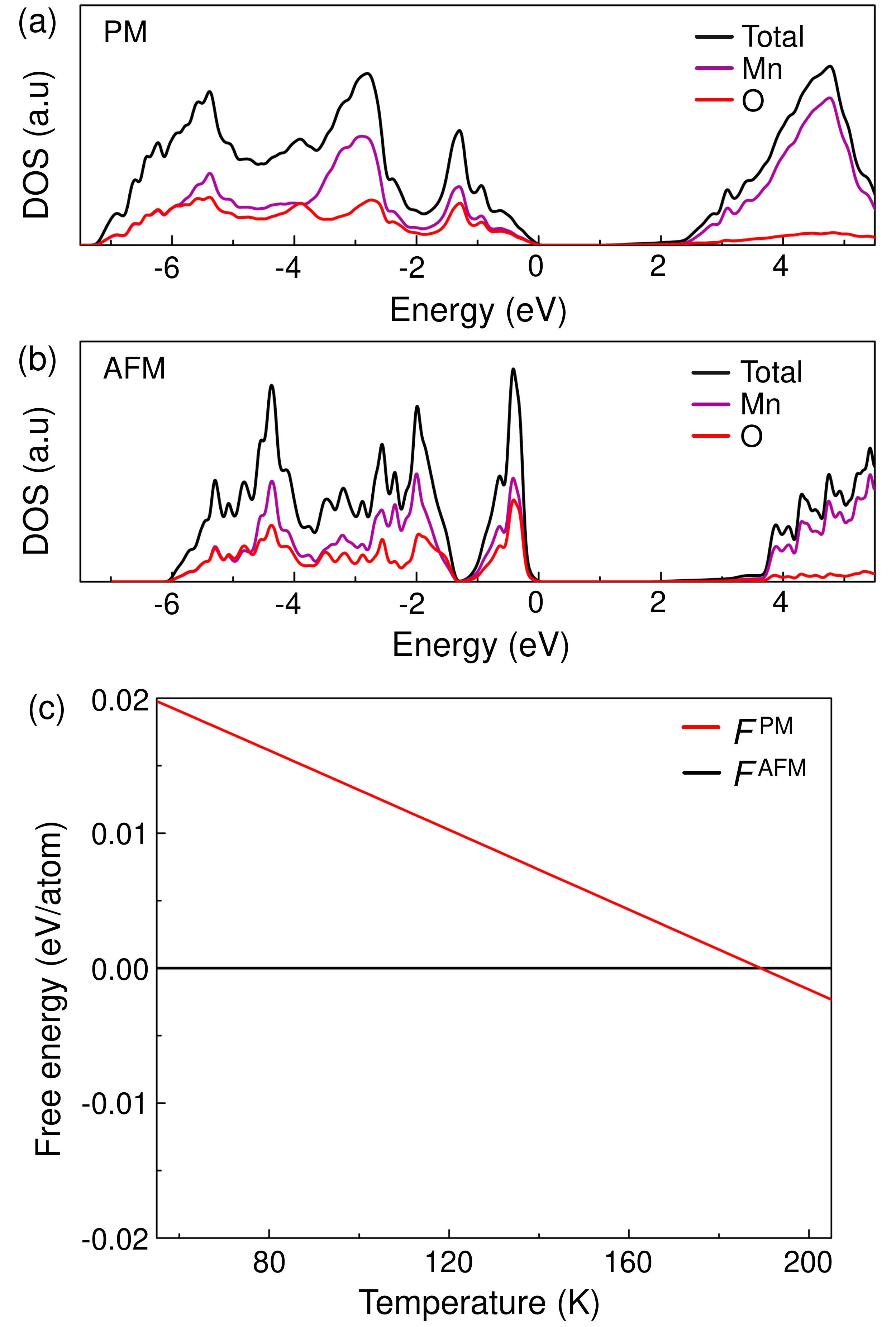}
\caption{(Color online)
Results of collinear calculations: DOSs of MnO in (a) paramagnetic and (b) antiferromagnetic states. (c) Per-atom Helmholtz free energy of paramagnetic (PM, red solid line) MnO relative to that of antiferromagnetic (AFM, black solid line) MnO, which is set to zero, obtained with $U_{\textrm{eff}}=4$~eV.
\label{Fig5}}
\end{figure}
%---------------------------------------------------------------------

The introduction of spin noncollinearity is a major feature of our NCMSM$+U$ approach compared to the conventional MSM. It is well known that, in most magnetic systems, the exchange interactions can be described by the bilinear exchange interactions. However, in some systems, the biquadratic exchange interactions cannot be ignored. The Heisenberg Hamiltonian that contains the biquadratic terms is generally expressed as 
\begin{align}
\mathcal{H} &=\sum_{i\neq j}J_{ij}\mathbf{S}_i\cdot\mathbf{S}_j+\sum_{i\neq j}K_{ij} (\mathbf{S}_i\cdot\mathbf{S}_j)^2 \nonumber \\
&\approx\sum_\alpha J_\alpha n_\alpha \langle\Phi_\alpha\rangle
+\sum_\alpha K_\alpha n_\alpha\langle\Psi_\alpha\rangle,
\label{equation6}
\end{align}
where $K_{ij}$ is the biquadratic exchange coupling constant. In the approximate Hamiltonian in the second row in Eq.~(\ref{equation6}), $\alpha$ is the index of coordination shell and $n_{\alpha}$ indicates the number of atoms in the $\alpha$-th coordination shell. Here, $\langle\Phi_\alpha\rangle$ and $\langle\Psi_\alpha\rangle$ are the bilinear and biquadratic spin-correlation functions, respectively, and $J_{\alpha}$ and $K_{\alpha}$ are their corresponding exchange coupling constants. In case of ideal paramagnetic disorder, $\langle\Phi_\alpha\rangle$ should be 0 in both collinear and noncollinear models. However, $\langle\Psi_\alpha\rangle$ has different values depending on whether the spin noncollinearity is allowed or not. $\langle\Psi_\alpha\rangle$ is estimated to be 1 in the collinear disorder model and 1/3 in the noncollinear counterpart. This implies that the NCMSM approach would yield significantly different computational results from the conventional MSM, unless the biquadratic interactions are sufficiently weak, or $K_{\alpha}$ is sufficiently small~\cite{PRB16}. Interestingly, MnO was directly studied by Anderson~\cite{PR4} and Orbach~\cite{PRL7}, who pointed out that its biquadratic superexchange interactions might have observable magnitudes. We, therefore, revised the MSM approach to include the spin noncollinearity for more accurate predictions of the high-temperature properties of MnO.

To examine the effects of the spin noncollinearity, we compared our computational results of paramagnetic MnO obtained by NCMSM$+U$ approach to those by MSM$+U$ approach. We designed all the computational details identical except for the spin noncollinearity. Figure~\ref{Fig5} shows the DOS of paramagnetic state (a) computed by the conventional MSM$+U$ approach and that of antiferromagnetic state (b). Even the conventional MSM$+U$ approach described the delocalized valence band, which is a significant feature of the paramagnetic MnO, as shown in Fig.~\ref{Fig2}(a). However, a remnant peak, which was not observed from the NCMSM$+U$, appears at the top of the valence band in the result of the MSM$+U$. Furthermore, we also estimated the N\'eel temperature based on the collinear calculations using the same method as described in Sec.~\ref{Results3}. Here, the N\'eel temperature was estimated to be $T_\mathrm{N}\approx189$~K, which is 60~\% higher than the experimental N\'eel temperature. Note that the N\'eel temperature estimated from our noncollinear calculation is almost equal to the experimental value. These results indicate that the biquadratic exchange interaction is significant in MnO, and, thus, the spin noncollinearity should be taken into account for accurate prediction of the physical properties of paramagnetic MnO.

It is noteworthy that the antiferromagnetic state of MnO does not require the spin noncollinearity even if the biquadratic exchange interactions are considered. MnO has a type-II antiferromagnetic ordering at low temperature, in which magnetic moments are all aligned in parallel. As a result, the spin-correlation functions of the collinear and noncollinear models are exactly the same: $\langle\Phi_\alpha\rangle$ represents a specific value depending on a coordination shell and $\langle\Psi_\alpha\rangle$ is always 1. We confirmed that the DOS of antiferromagnetic MnO based on the collinear model shown in Fig.~\ref{Fig5}(b) is equivalent to that based on the noncollinear counterpart shown in Fig.~\ref{Fig1}(b).

% Conclusions
\section{Conclusions}
\label{Conclusions}

We employed the noncollinear magnetic sampling method (NCMSM) with DFT$+U$ (NCMSM$+U$) to calculate the paramagnetic Mott insulating state of MnO and investigated its room-temperature properties. The NCMSM$+U$ approach accurately predicts the physical properties of paramagnetic Mott insulator MnO at reasonable computational cost. Specifically, we estimated the electronic profile, distribution of local magnetic moments, superexchange coupling constants, and N\'eel temperature. Furthermore, our work demonstrated that the inclusion of spin noncollinearity plays a crucial role in accurate description of paramagnetic Mott insulators where superexchange interactions exist. It significantly affects the detailed electronic profile and precise energetics of paramagnetic MnO. The NCMSM$+U$ approach is of value in terms of scalability to other supercell calculations as compared to KKR-CPA or DFT+DMFT. We, therefore, expect our results to help providing a basis for the rational design of energy materials using paramagnetic Mott insulators. It should be noted that the NCMSM$+U$ approach cannot be used for paramagnetic metals with fast dynamic charge correlations.

%Acknowledgements
\section*{Acknowledgements}
We acknowledge financial support from the Korean government through National Research Foundation (2017R1A2B3011629, 2019R1A2C1005417), and from the Ministry of Trade, Industry \& Energy (MOTIE) of Korea (Project No. 10080625). Some portion of our computational work was done using the resources of the KISTI Supercomputing Center (KSC-2018-C2-0033 and KSC-2018-CHA005).

%+++++++++++++++++++++++++++++++++++++++++++++++++++++++++++++++++++++
%References

\end{document}